\documentclass[sn-mathphys-num,iicol]{sn-jnl}

\usepackage{graphicx}%
\usepackage{multirow}%
\usepackage{amsmath,amssymb,amsfonts}%
\usepackage{amsthm}%
\usepackage{mathrsfs}%
\usepackage[title]{appendix}%
\usepackage{xcolor}%
\usepackage{textcomp}%
\usepackage{manyfoot}%
\usepackage{booktabs}%
\usepackage{algorithm}%
\usepackage{algorithmicx}%
\usepackage{algpseudocode}%
\usepackage{listings}%
\usepackage{calrsfs}%
\DeclareMathAlphabet{\pazocal}{OMS}{zplm}{m}{n}
\usepackage{dsfont}
\usepackage{lmodern}
\usepackage{anyfontsize}

\begin{document}

\title[Article Title]{Probing late-time annihilations of oscillating asymmetric dark matter via rotation curves of galaxies}


\author*[1]{\fnm{Júlia} \sur{G. Mamprim}}\email{juliagouveamamprim@usp.br}

\author[1,2]{\fnm{Guillermo} \sur{Gambini}}

\author[1]{\fnm{Luan} \sur{B. Arbeletche}}

\author[1]{\fnm{Marcos} \sur{Olegario}}

\author[1]{\fnm{Vitor} \sur{de Souza}}

\affil*[1]{\orgdiv{Instituto de Física de São Carlos}, \orgname{Universidade de São Paulo}, \orgaddress{\street{Av. Trabalhador São-carlense 400}, \city{São Carlos}, \postcode{13566-590}, \state{São Paulo}, \country{Brazil}}}

\affil[2]{\orgdiv{Facultad de Ciencia}, \orgname{Universidad Nacional de Ingenieria (UNI)}, \orgaddress{\street{Av. Túpac Amaru 210}, \city{Lima},  \postcode{15032}, \country{Perú}}}


\abstract{In this paper, we explore the Oscillating Asymmetric Dark Matter (OADM)
model to address the core-cusp problem, aiming to resolve the discrepancy between the
predictions of the $\Lambda\rm{CDM}$ cosmological model and the observed dark matter profiles
in dwarf spheroidal galaxies. The reactivation of dark matter annihilation during the
structure formation epoch is possible if there is a small Majorana mass term that breaks
the conservation of dark matter particle number, leading to oscillations between dark
matter and its antiparticle. We analyzed the effects of the annihilation mechanism in the
galaxy rotation curves of the SPARC and LITTLE THINGS catalogs. We searched for
the characteristics of the OADM model which best describes the data. Our results show
that the OADM model can successfully turn originally cusp-type halos into core-type
ones according to our data sample.}




\maketitle

\section{Introduction}\label{sec:intro}

The nature of Dark Matter (DM) remains one of the most fundamental unsolved questions in physics. Numerous observations across different scales provide strong evidence for the presence of DM, including galaxy rotation curves \citep{Sofue:2000jx}, the dynamics of galaxy clusters~\citep{Clowe:2006eq}, fluctuations in the cosmic microwave background~\citep{Planck:2018vyg}, baryon acoustic oscillations~\citep{SDSS:2005xqv}, among others. However, the exact nature of Dark Matter remains unclear due to the lack of results from both direct detection experiments~\citep{LUX:2016ggv,Arcadi:2017kky} and indirect detection studies~\citep{HESS:2018cbt,HESS:2016mib,MAGIC:2016xys,Fermi-LAT:2019lyf}.

In 1970, Rubin and Ford~\citep{1970ApJ...159..379R} conducted the first precise measurement of the rotation curve of the Andromeda galaxy (M31) and found that the velocity profile was almost flat at large radii. This unexpected result suggested the abundance of dark matter within M31 and was later corroborated by observations of dozens of other galaxies. Despite providing intuitive evidence for the existence of dark matter, it is currently a challenge to conceal the observed rotation curves to theoretical aspects of the dark matter, in particular, their density profiles.

Several models attempt to describe the distribution of dark matter
in galaxies, including the Navarro-Frenk-White (NFW)~\citep{NFW}, Moore~\citep{Moore:1997sg} and Einasto~\citep{1965TrAlm...5...87E} halo density profiles, among others. These models were developed using N-body simulations based on the \(\Lambda\)CDM cosmological framework, without accounting for baryonic feedback, and suggest a steep DM distribution in the central region of galaxies. However, observations of late-type dwarf galaxies, which are dominated by dark matter, reveal a roughly constant DM density at small radii. This discrepancy is known as the “core/cusp problem” ~\citep{de_Blok_2009, Salucci:2018hqu}. Including baryonic feedback in N-body simulations could potentially flatten the central cusps of halos in massive galaxies. However, it remains uncertain whether this solution is effective in the lowest mass galaxies, where such discrepancies are observed~\citep{Weinberg:2013aya}, or in systems like dwarf spheroidal galaxies, which have relatively low baryonic content.

Self-Interacting Dark Matter (SIDM) models present another potential solution to the core/cusp problem. Early SIDM simulations required scattering cross sections around \(\sigma/m_\chi \sim (0.4 - 5) \, \text{cm}^2/\text{g}\)~\citep{Spergel:1999mh,Tulin:2017ara}. For small structures like dwarf spheroidal galaxies, removing dark matter cusps using SIDM models requires cross sections \(> 1 \, \text{cm}^2/\text{g}\) at \(v_{\chi} \sim 10 \, \text{km}/\text{s}\)~\citep{Eckert:2022qia} while most stringent constraints from merging galaxy clusters suggest \(\sigma/m_\chi < 0.1 \, \text{cm}^2/\text{g}\) \citep{Markevitch:2003at}. SIDM models with velocity-dependent scattering cross-section that decreases with increasing velocity~\citep{Kaplinghat:2015aga} could still make SIDM a viable solution to the core-cusp problem. 

Our work directly builds off of reference \cite{Cline:2020gon}, where it was shown, for representative benchmarks, that late-time dark matter annihilations in galactic structures could be responsible for erasure of the cusps in galactic dark matter density profiles while being allowed by CMB constraints \citep{Poulin:2016nat,Bringmann:2018jpr}. We perform a wider search of parameter space by fitting 37 rotation curves of galaxies using observational data from SPARC~\citep{Lelli:2016zqa} and LITTLE THINGS~\citep{Hunter:2012un}, which have been selected for their high-quality rotation curves and high mass-to-light ratios. Our results demonstrate that the preferred window for the parameter space of fermionic OADM annihilating into a lighter scalar is the mass range [100-1000] MeV and dark fine structure constant $\alpha'$ around $0.1$. Moreover, we find $\sigma/m_\chi \approx 0.06$ cm$^2/$g for the annihilation cross section is prefered by the $\chi^2$ minimization and it presents very little scatter. 

In Section~\ref{sec:oadm}, we define the OADM model framework. Section~\ref{sec:density} describes how density profiles evolve from cusp-type to core-type within our model. Section~\ref{sec:rotation} discusses the observational data and the methodology for fitting rotation velocity curves. Section~\ref{sec:results} presents the results of the combined fits. Finally, Section~\ref{sec:conclusion} summarizes the findings and concludes the work.

\section{Oscillating asymmetric dark matter formalism}\label{sec:oadm}

SWe consider a dark matter field \(\chi\) that is a quasi-Dirac fermion with mass \(m_\chi\). In this model, the DM couples to a complex scalar \(\Phi = \phi + ia\), with effective Lagrangian
\begin{equation}
    \pazocal{L} \supset -\frac{1}{2}m^2_{\phi} \phi^2 - \frac{1}{2}m^2_aa^2 - g^{\prime} \Bar{\chi}(\phi + ia\gamma_5)\chi, 
\end{equation}
 where \(g^{\prime}\) denotes the coupling between the DM particle and the scalar boson, with an associated dark fine-structure constant \(\alpha^{\prime} = g^{\prime 2}/4\pi\). Dark matter freeze-out and late-time
depletion are both allowed by the \(s\)-wave process \(\chi\Bar{\chi} \rightarrow a \phi\) (which, unlike the \(p\)-wave channels \(\chi \Bar{\chi} \rightarrow \phi \phi\) and \(\chi \Bar{\chi} \rightarrow aa\), is not suppressed at low velocities).
 
The mass term that permits DM-number violation is

\begin{equation}
    \pazocal{L}_m = \frac{1}{2} \delta m (\Bar{\chi}\chi^c + \text{H.c.}),
\end{equation}
 with the Majorana mass \(\delta m\) determining the timescale on which annihilations recouple after the freeze-out. The parameter \(\delta m \) must be small enough so that the oscillations between DM and its antiparticle have not yet started at the time of freeze-out. For a flavor-blind
interaction, where scattering does not play a role as this type of interaction does not distinguish between $\chi$ and $\bar{\chi}$, annihilations start when oscillations repopulate the symmetric dark matter density, i.e., for \(t \gtrsim \delta m^{-1}\) \citep{Kaplan:2009ag,Tulin:2012re}. Moreover, the oscillations should start before structure formation, \(t_\text{s} \sim 0.1 ~\text{Gyr}\) for annihilations to recouple during that epoch. However, if the annihilations are reactivated too early, the changes on DM relic density would be ruled out by CMB constraints \citep{Poulin:2016nat,Bringmann:2018jpr}. Taking the fractional change in the DM number density as \(\delta_\eta \sim 3\%\) (limited by the change in the dark matter density after the formation of the CMB), and considering \(m_\chi \sim 100\) MeV, an approximate window of values for \(\delta m\) can be set \citep{Cline:2020gon}:

\begin{equation}
    \frac{1}{t_s} \lesssim \delta m \lesssim 3 \times 10^{-30} \text{eV}.
    \label{eq:dm}
\end{equation}
Eur. Phys. J. C manuscript No.
Above \(m_\chi \sim 1 \rm{GeV}\), the theory starts to be strongly coupled \citep{Cline:2020gon}, so we restrain our analysis to this maximum mass value.

The annihilation cross section at threshold, \textit{i.e.} the minimum energy at which the annihilation process can occur, reads
\begin{equation}
    \langle \sigma v \rangle_a = \frac{\pi \alpha^{\prime 2}}{m^2_\chi}\left(1 - \frac{m_\phi^2}{4m_\chi^2}\right),
    \label{eq:xsec}Eur. Phys. J. C manuscript No.
\end{equation}
where we have assumed, for simplicity, the following hierarchy  \(m_a \ll m_\phi\), and neglected the suppressed channels \(\chi \Bar{\chi} \rightarrow \phi \phi\) and \(\chi \Bar{\chi} \rightarrow aa\). 

Even though $\chi \chi$ elastic scatterings are irrelevant for the reactivation of dark matter annihilations at late times when the interaction is flavor-blind, as in this model, we consider values in the parameter space for which DM annihilations are dominant over scattering self-interactions which, otherwise, could change the DM density profiles in the center of galaxies as in SIDM models.

When DM oscillations are present, the difference between particle and antiparticle varies over time. For interactions that don't distinguish between a dark matter particle and its antiparticle, the evolution of the DM and anti-DM number densities, encoded in the 2 by 2 matrix $n$, is best described by the Boltzmann equation \citep{Tulin:2012re}

\begin{multline}
        \Dot{n} + 3Hn = -i[H_0,n] - \langle\sigma v\rangle_a \times \\ \times \left[ 
    \begin{pmatrix}
        \text{det}^{\prime} n & (\text{Tr}n)n_{12} \\
        (\text{Tr}n)n_{21} & \text{det}^{\prime}n
    \end{pmatrix}
    - n^2_{\rm{eq}} \mathds{1} \right],
    \label{eq: Boltzmann}
\end{multline}
where \(H\) is the Hubble parameter, \(n_{\rm{eq}}\) is the equilibrium number density, and \(\rm{det}^\prime \equiv n_{11}n_{22} + n_{12}n_{21}\). As an approximation, we considered the DM density to be spatially homogeneous and isotropic for an already formed DM halo, simplifying the terms that come from the Liouville operator in the Boltzmann equation \citep{Kolb:1990vq}. In that sense, the diffusion of dark matter particles in the system is not considered by this approach. However, N-body simulations show close results to the Boltzmann method \citep{Cline:2020gon}, but this analysis is outside the scope of this work.

Ref. \cite{Cline:2020gon} also describes another DM model, with a vector mediator and flavor-sensitive interactions. In contrast to the scalar model, the Boltzmann equation for the flavor-sensitive case presents a dependence on the DM velocity, which could lead to different levels of change in the halo density profiles depending on the size of the galactic system considered. Since we are performing a combined analysis based on data from two distinct rotation curves catalogs, with galaxies that vary in extension, it is a reasonable choice to probe only the scalar model's parameters trough this method.

\section{A model for halo density profile evolution with OADM}\label{sec:density}

To test the effect of the flavor-blind OADM model on galaxy rotation curves, we consider an initial NFW-shaped dark matter halo, already formed at time \(t_0\)~\citep{NFW}: 

\begin{equation}
    \rho_{\mathrm{NFW}, t_0} = \frac{\rho_s}{( r/r_s)(1+r/r_s)^2},
    \label{eq: NFW}
\end{equation}
where \(r_s\) is the scale radius of the system, and \(\rho_s\) is the scale density \(\rho_s = \rho_c \delta_c\), given as a function of the critical density of the Universe \(\rho_c\) and an over density parameter \(\delta_c = (200/3)(c^3_{200}/f(c_{200}))\). Here, \(c_{200}\) is the concentration parameter, defined from the Virial radius \(r_{200}\) as \(c_{200} = r_{200}/r_s\), and \(f(c_{200})\) corresponds to the function \(f(c_{200}) = \text{ln}(1 + c_{200}) - c_{200}/(1 + c_{200} )\).

Before the oscillations have any effect on the collapsed halo environment, the initial condition at each position \(r\) for the density matrix is 

\begin{equation}
    n_{ij}(r;t_0)=\frac{\rho_{{\rm NFW},t_0}(r)}{m_\chi} \delta_{i1}\delta_{j1},
    \label{eq: initialdensity}
\end{equation}
which represents a pure dark matter \(\chi\) state.

For a galactic halo, where the expansion of the universe can be safely neglected, equation \ref{eq: Boltzmann} simplifies by dropping the \(3Hn\) term and the equilibrium function since \(n_{\rm{eq}} \to 0\) at late times. Furthermore, since NFW profiles vary orders of magnitude from the inner part of a galaxy to its outskirts, we can assume the products of the annihilations can escape from the galactic system without interacting again.

Under these assumptions, we evolve the initial density matrix \ref{eq: initialdensity} at each radial position \(r\) from \(t_0 \sim 0.1 \rm{Gyr}\) until \(t_f \sim 10 \rm{Gyr}\) \citep{Kaplinghat:2015aga,Cline:2020gon}, leading to a modified density profile \(\rho_{\chi}(t_f) = (n_{11}(t_f) + n_{22}(t_f)) m_\chi\) that we compare to observational data.

\section{Fitting measured galaxy rotation curves}\label{sec:rotation}

To probe the parameter space of the aforementioned OADM model, we utilized observed rotation curves of 37 galaxies from two different database catalogs, namely SPARC (Spitzer Photometry \& Accurate Rotation Curves) \citep{Lelli:2016zqa} and  LITTLE THINGS (Local Irregulars That Trace Luminosity Extremes, The H\scriptsize{I} \normalsize Nearby Galaxy Survey) \citep{Hunter:2012un}. 

SPARC is a sample of 175 late-type galaxies with high-quality H\scriptsize{I}\normalsize\(/\rm{H}\alpha\) rotation curves and near-infrared Spitzer photometry. The morphologies vary among spirals and irregulars, presenting a wide range in stellar masses, surface brightnesses, and gas fractions. The mass distribution for stellar disk and bulge is traced with Spitzer photometry at \(3.6 \mu\rm{m}\), and the mass contribution of the gas is derived from
the observed H\scriptsize{I} \normalsize surface density profile, scaled up to include Helium \citep{Li:2020iib}.

A high baryonic component in the galactic environment could lead to effects that we did not consider in our model, such as baryonic feedback and more expressive non-circular contributions to the galaxy motion. Therefore, we made a conservative cut in the SPARC sample, selecting only the 18 galaxies that have a mass-to-light ratio larger than \(4 M_\odot/L_\odot\). All of the selected galaxies do not present a clear bulge region (a dense, spherical region of stars located at the center of many spiral galaxies), and are described only by gas, disk and halo components.  

The LITTLE THINGS sample includes 37 dwarf irregular (dIm) and 4 Blue Compact Dwarf (BCD) galaxies. A subset of 26 galaxies that show regular rotation pattern in their two-dimensional H\scriptsize{I}\normalsize\,velocity fields had their rotation curves and mass models derived in Ref. \cite{Oh:2015xoa} from the high-resolution H\scriptsize{I}\normalsize\,observations, complemented with Spitzer \(3.6 \mu\rm{m}\) measurements, as well as \(\rm{H}\alpha\),  optical  \(U, B, V\) and near-infrared images.
Among those 26, we excluded three galaxies (DDO50, IC10, IC1613) that presented high baryonic gas content and other three galaxies (DDO46, DDO101, NGC3738) that had an isothermal halo profile scale density \(\rho_0 \gtrsim 500 \times 10^{-3} M_\odot \rm{pc^{-3}}\) \citep{Oh:2015xoa}, because a very high initial scale density could lead annihitilations to be too efficient, with the total density profile vanishing at the timescales (\(\sim 10 \rm{Gyr}\)) we are working with. It is important to emphasize that the model's mechanism is effective for the systems in question but only within shorter evolution intervals (\(\sim 5 \rm{Gyr}\)). This limitation may suggest that these halos have not yet reached equilibrium, with their concentration parameters still evolving toward their final values. To avoid introducing the evolution time of the systems as an additional parameter in the fit, we opted to exclude these outlier galaxies from the analysis. Therefore, after the quality cuts, we were left with 20 galaxies from the LITTLE THINGS catalog. 

The rotation curves from both samples were compared to the rotation velocity curves obtained by integrating our modified density profile \(\rho_{\mathrm{\chi}, t_f}\),

\begin{equation}
       V_{{\chi}}^2(r) =  \frac{GM(r)}{r} = \frac{4\pi G}{r}\int_0^r r'^2 \rho_{{\chi}, t_f}(r')dr' 
       \label{velocidade}.
    \end{equation}

The contribution of each galactic component is accounted by the relation 

\begin{equation}
    V^2_{\text{tot}} = V^2_{\chi} + \Upsilon_\star V^2_\star + V^2_\text{gas},
\end{equation}
where \(V_\text{tot}\) is the total velocity expected after evolving the initial density profile, \(V_{\chi}\) is the dark matter halo component, \(V_\star\) is the stellar disk contribution, \(V_\text{gas}\) comprise the gas fraction, and \(\Upsilon_\star\) is the stellar mass-to-light ratio. While the mass of the gas component can be directly estimated from H\scriptsize{I}\normalsize\,observations, the stellar mass in galaxies is
strongly dependent on the assumed \(\Upsilon_\star\), which is usually responsible for the largest uncertainties when converting the luminosity profile to the mass density profile \citep{Lelli:2016zqa}. For SPARC and LITTLE THINGS galaxies, respectivelly, we took the \(\Upsilon_\star\) values as discussed in Ref. \cite{Lelli:2016zqa} and Ref. \cite{Oh:2015xoa}.

The free parameters of our model are adjusted to minimize the $\chi^2$ statistic, defined as

\begin{equation}
\chi^2 = \sum_{r_i}\frac{[V_{\text{obs}}(r_i) - V_{\text{tot}}(r_i)]^2}{\sigma^2_i},
\end{equation}
where \(V_{\text{obs}}\) is the observed rotation velocity, \(\sigma_i\) is
the observational uncertainty and \(V_{\text{tot}}\) is the total rotation velocity expected after evolving the initial density profile. We perform a grid-search over the \(\alpha^\prime \times m_\chi\) parameter space for 50 linearly spaced steps in the ranges \([0.01,0.3]\) and \([100.0, 1000.0]\) MeV, respectively. For each pair $(\alpha^\prime, m_\chi)$, we fit the observations to a NFW-profile, with two free parameters: the halo scale radius \(r_s\) and the scale density \(\rho_s\).

\begin{figure}
    \centering
    \includegraphics[width=0.45\textwidth]{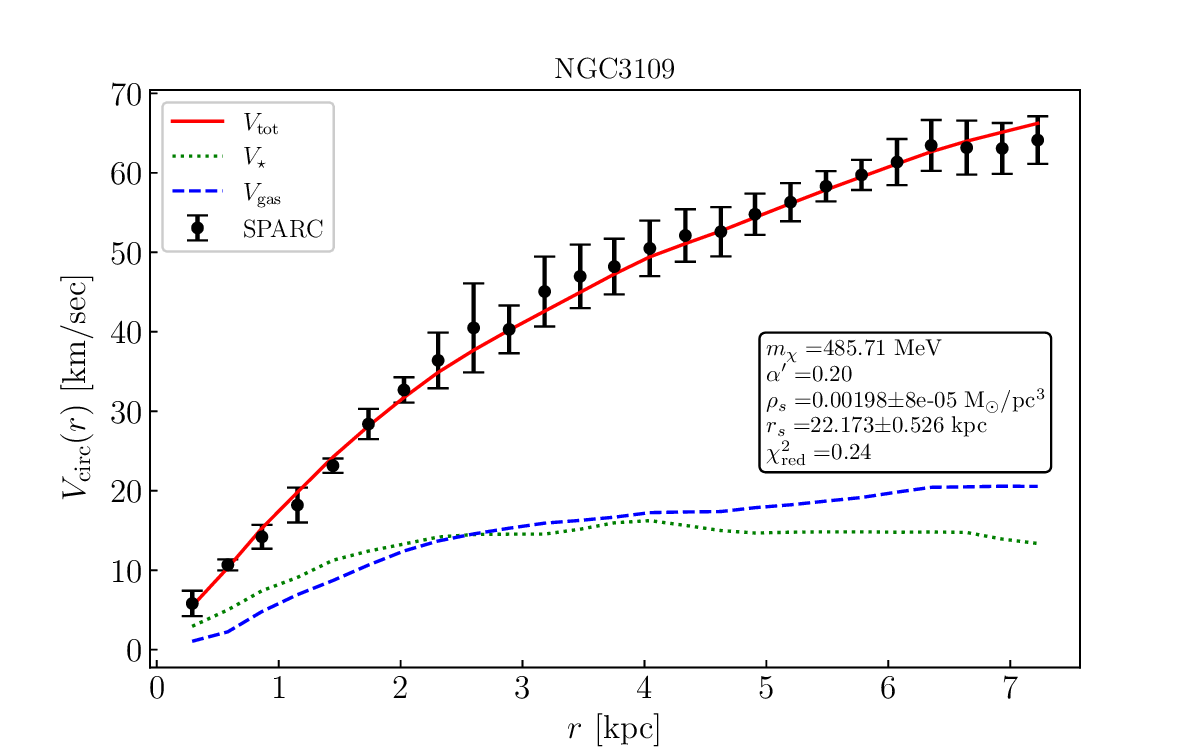}
    \caption{Rotational velocity as a function of the distance from the center. Box shows the result of the NFW profile with OADM model for the galaxy NGC3109, with fixed values for \(\delta m = 1.0\times10^{-31}\) and \(m_\phi = 0.7m_\chi\). The black dots show the observational rotation velocity data available in the SPARC catalog. The green dotted line is the gas component, and the blue dashed line is the stellar disk component, both given by SPARC. The red line is the result of the fit for \(V_\text{tot}\). The reduced \(\chi^2\) is given by \(\chi^2_\text{red} = \frac{\chi^2}{n - m}\), where \(n\) equals the number of observations, and \(m\) is the number of fitted parameters.}
    \label{fig:NGC3109}
\end{figure}

\begin{figure}
    \centering
    \includegraphics[width=0.45\textwidth]{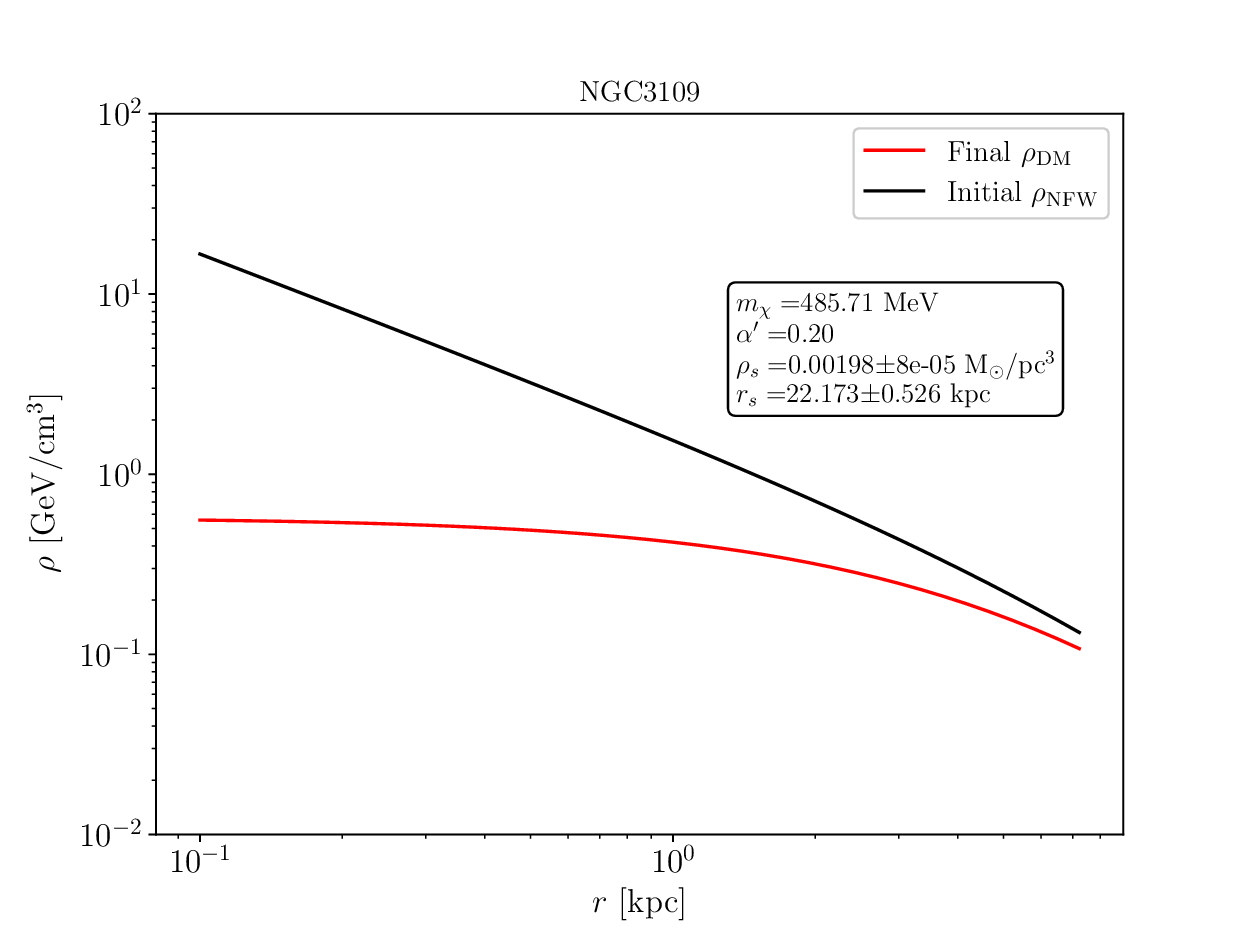}
    \caption{Dark matter density as a function of the distance from center for the galaxy NGC3109, with fixed values for \(\delta m = 1.0\times10^{-31}\) and \(m_\phi = 0.7m_\chi\). The black line is the initial NFW profile, with \(\rho_s\) and \(r_s\) as shown in the figure. The red line shows 
    the final density profile, affected by the annihilations in the OADM model.} 
    \label{fig:NGC3109density}
\end{figure}

\begin{figure}
    \centering
    \includegraphics[width=0.45\textwidth]{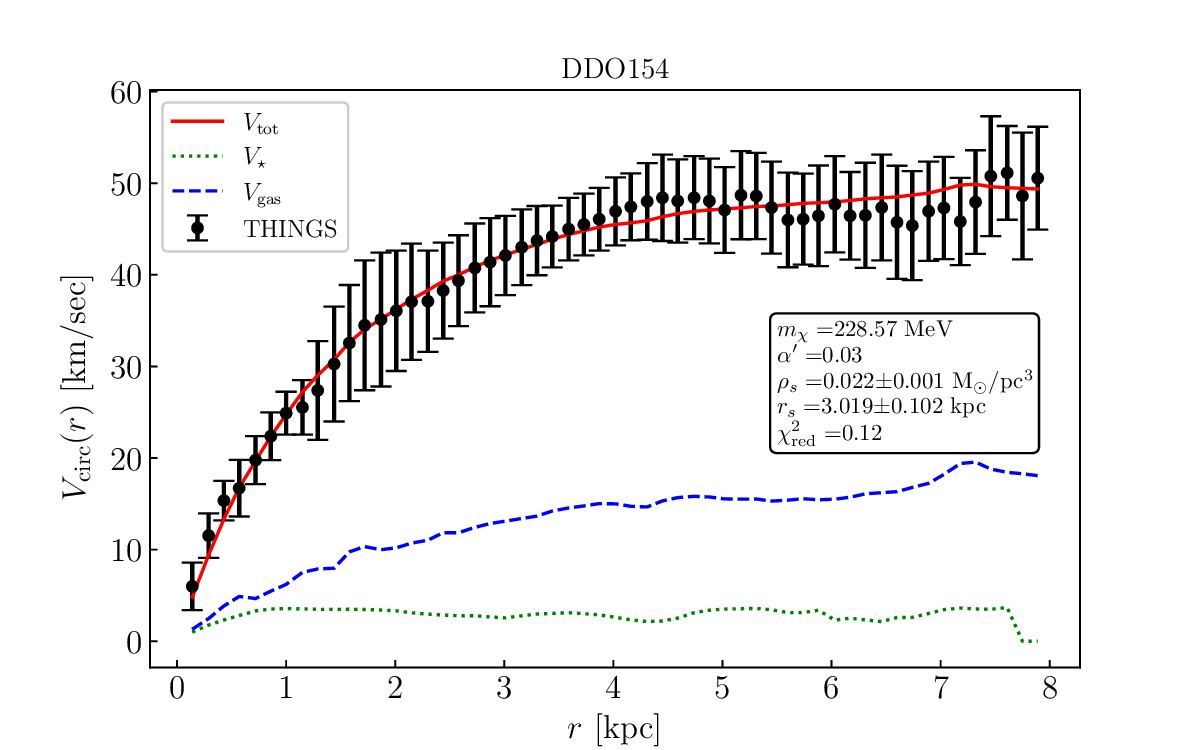}
    \caption{Rotational velocity as a function of the distance from the center. Box shows the result of the NFW profile with OADM model for the galaxy DDO154, with fixed values for \(\delta m = 1.0\times10^{-31}\) and \(m_\phi = 0.7m_\chi\). The black dots show the observational rotation velocity data available in the LITTLE THINGS catalogue. The green dotted line is the gas component, and the blue dashed line is the stellar disk component, both given by LITTLE THINGS. The red line is the result of the fit for \(V_\text{tot}\). The reduced \(\chi^2\) is given by \(\chi^2_\text{red} = \frac{\chi^2}{n - m}\), where \(n\) equals the number of observations, and \(m\) is the number of fitted parameters.}
    \label{fig:DDO154}
\end{figure}

\begin{figure}
    \centering
    \includegraphics[width=0.45\textwidth]{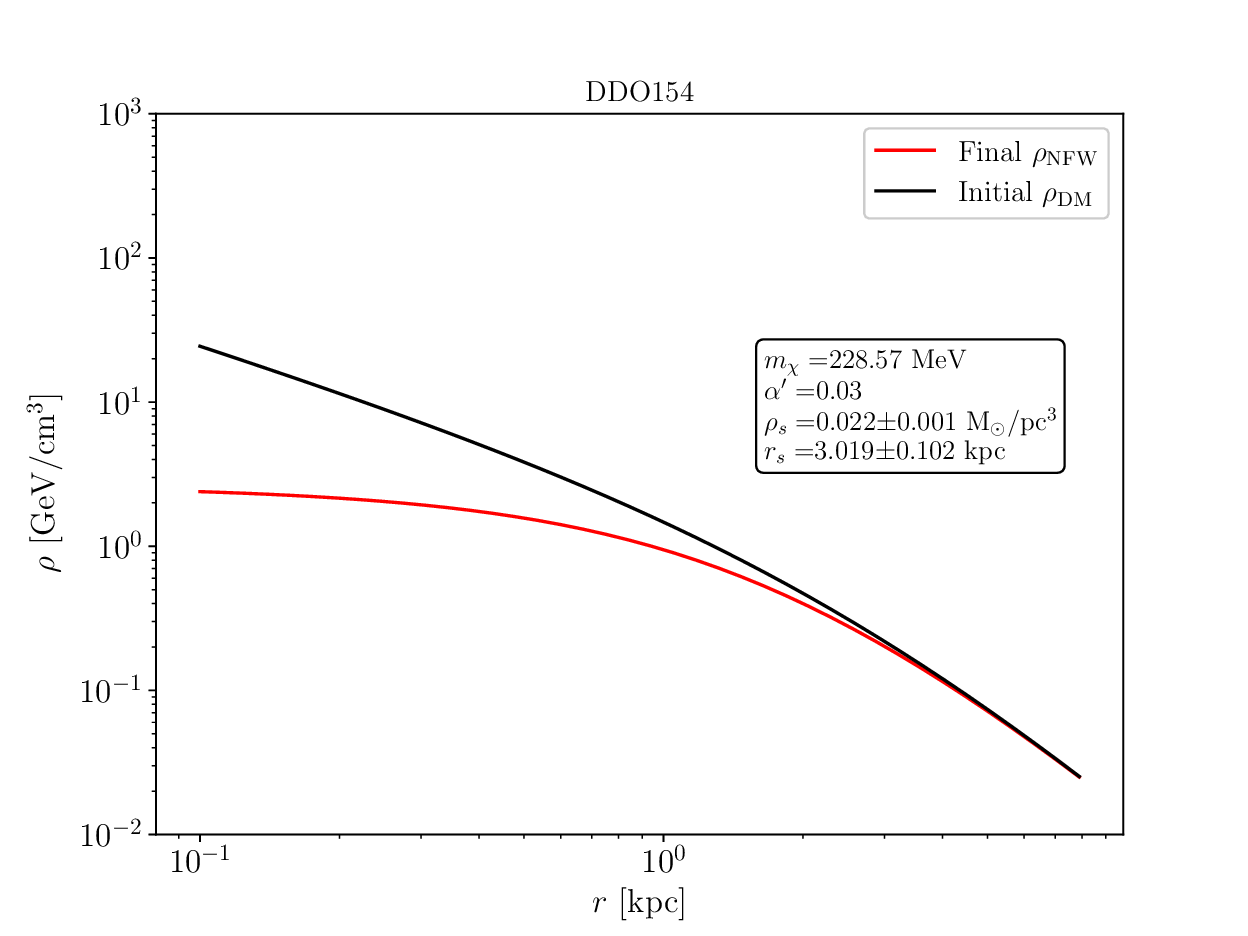}
    \caption{Dark matter density as a function of the distance from center for the galaxy DDO154, with fixed values for \(\delta m = 1.0\times10^{-31}\) and \(m_\phi = 0.7m_\chi\). The black line is the initial NFW profile, with \(\rho_s\) and \(r_s\) as shown in the figure. The red line shows 
    the final density profile, affected by the annihilations.} 
    \label{fig:DDO154density}
\end{figure}

Figures~\ref{fig:NGC3109} and~\ref{fig:DDO154} show the result of the fit for NGC3109 in the SPARC catalog and for DDO154 in the LITTLE THINGS catalog, respectively, as an example. Figures~\ref{fig:NGC3109density} and~\ref{fig:DDO154density} show the corresponding evolution of the halo density profile of these galaxies. On Table \ref{table:Table1}, we present the best-fit parameters corresponding to the minimum \(\chi^2_{\rm{red}}\) in the \(\alpha^\prime \times m_\chi\) grid, where \(\chi^2_\text{red} = \frac{\chi^2}{n - m}\) is the reduced \(\chi^2\), with \(n\) being the number of observations, and \(m\) the number of fitted parameters. We see that, in the grid-search, a good fit was found for all the galaxies in our sample. 

The first two parameters in our model, \(r_s\) and \(\rho_s\), come from the initial NFW halo density profile, and are expected to differ from one galaxy to another. The parameters \(m_\chi\) and \(\alpha^\prime\), on the other hand, relate to the OADM model and should, in principle, be the same for all galaxies. By fitting these parameters individually for each galaxy, however, we did not verify this last statement. Therefore, in order to explore the pair of parameters \(m_\chi\) and \(\alpha^\prime\) that best describe our ensembles of galaxies, a combined approach is necessary, which is presented in the next section.

\section{Results of the fit for the catalogs of galaxies}
\label{sec:results}

\begin{figure}
    \centering
    \includegraphics[width=0.5\textwidth]{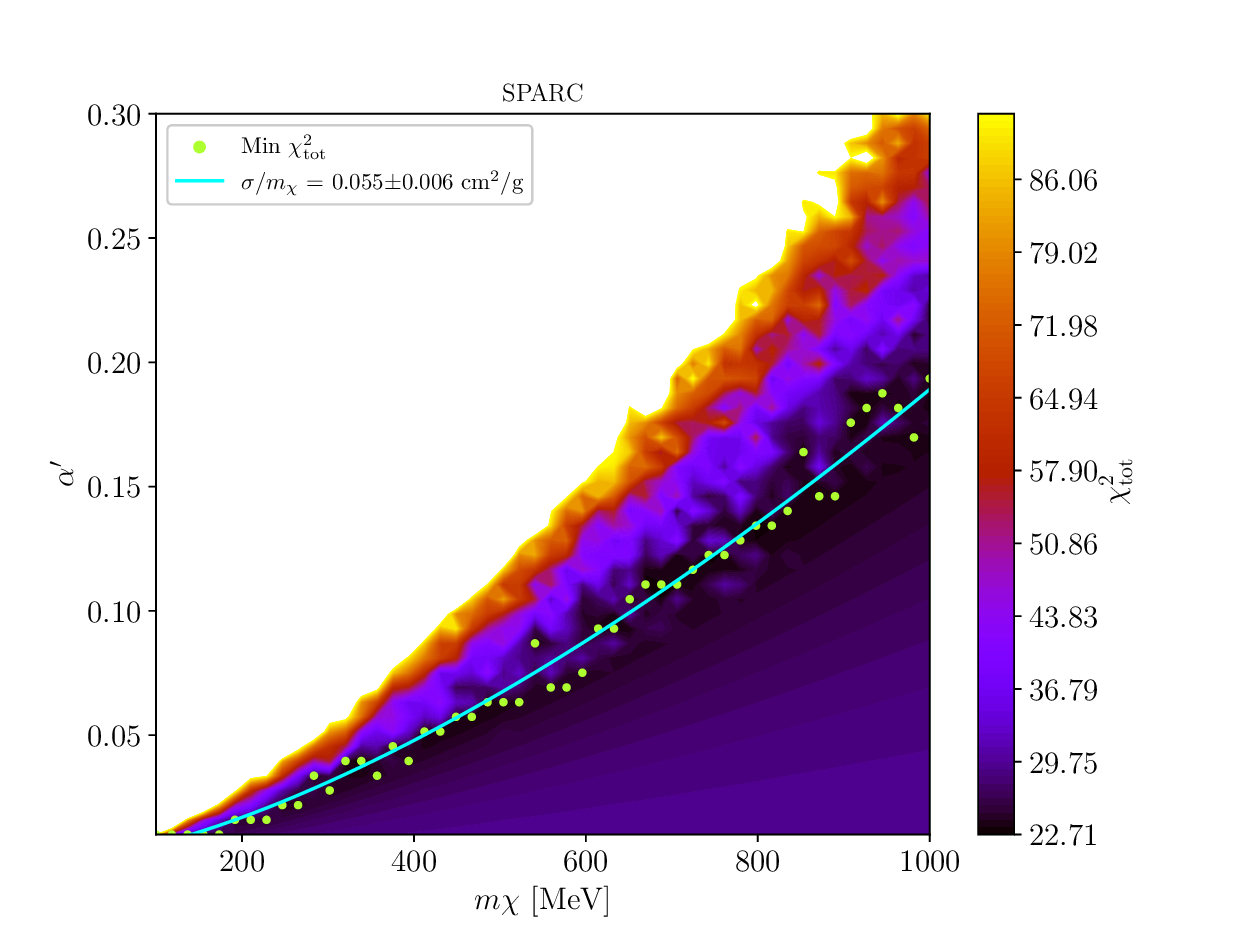}
    \caption{Value of the \(\chi^2_{\rm{tot}}\) in the \(\alpha^\prime \times m_\chi\) parameter space for the 18 galaxies from SPARC. The green markers correspond to the minimum of \(\chi^2_{\rm{tot}}\) at discrete values of \(\alpha^\prime\) and \(m_\chi\). The blue line shows the fit with the function \ref{eq:function_alpha_m}.}
    \label{fig:SPARC_combined}
\end{figure}

\begin{figure}
    \centering
    \includegraphics[width=0.5\textwidth]{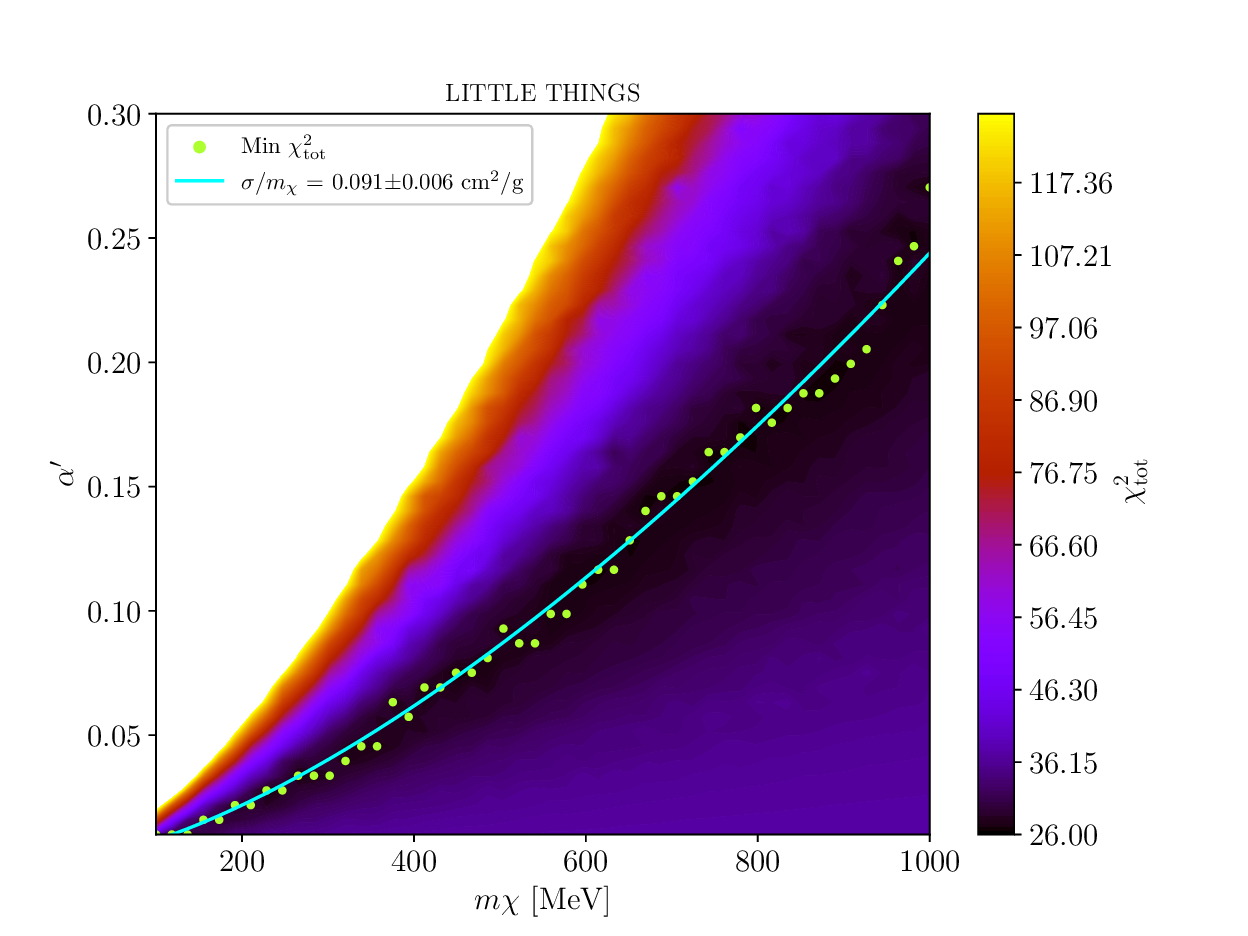}
    \caption{Value of the \(\chi^2_{\rm{tot}}\) in the \(\alpha^\prime \times m_\chi\) parameter space for the 20 galaxies from LITTLE THINGS. The green markers correspond to the minimum of \(\chi^2_{\rm{tot}}\) at discrete values of \(\alpha^\prime\) and \(m_\chi\). The blue line shows the fit with the function \ref{eq:function_alpha_m}..}
    \label{fig:THINGS_combined}
\end{figure}

\begin{figure}
    \centering
    \includegraphics[width=0.5\textwidth]{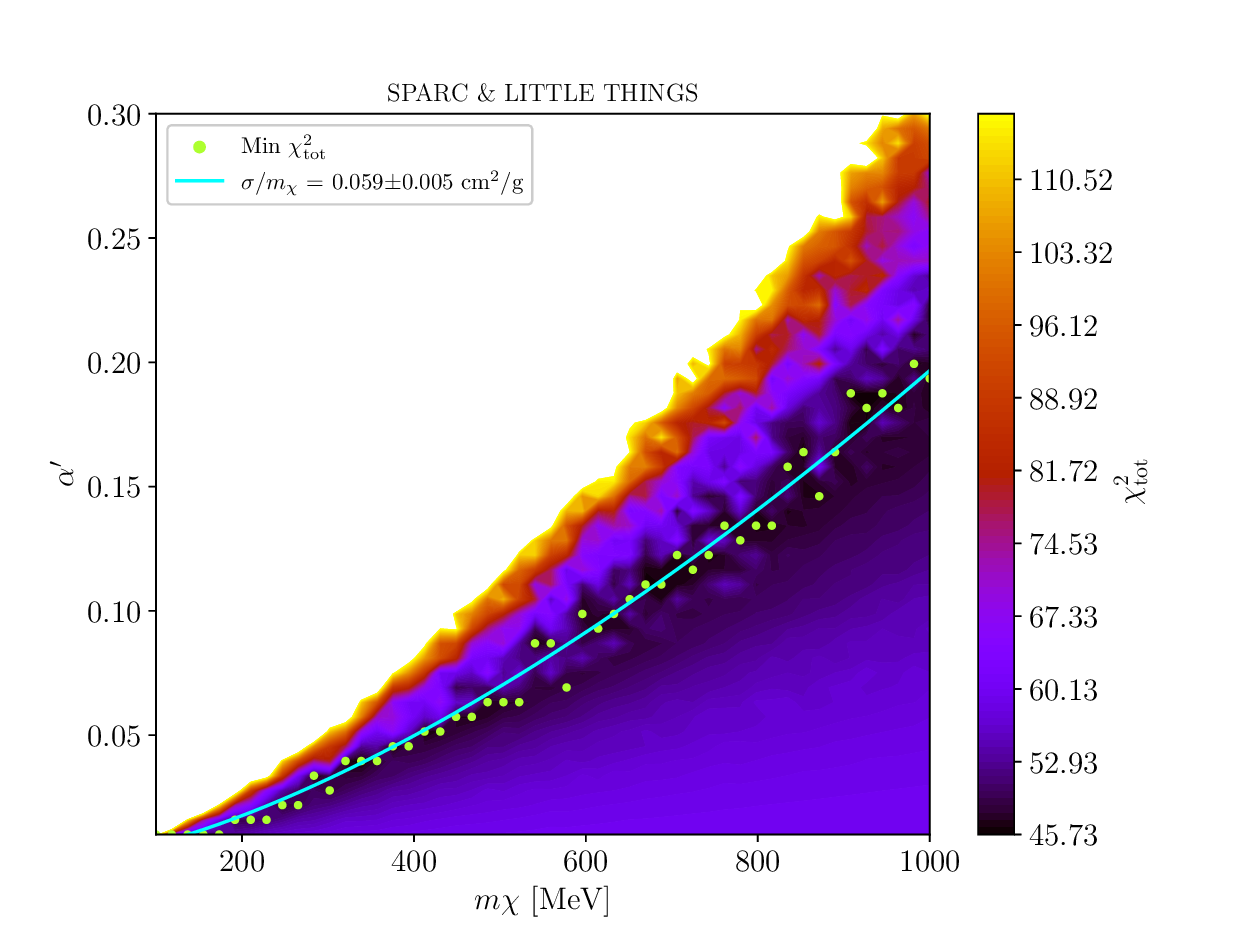}
    \caption{Value of the \(\chi^2_{\rm{tot}}\) in the \(\alpha^\prime \times m_\chi\) parameter space for 37 galaxies from SPARC and LITTLE THINGS togheter. Galaxy DDO154 appears in both catalogues, and we chose to use LITTLE THINGS data since it presented more data points. The green markers correspond to the minimum of \(\chi^2_{\rm{tot}}\) at discrete values of \(\alpha^\prime\) and \(m_\chi\). The blue line shows the fit with the function \ref{eq:function_alpha_m}.}
    \label{fig:BOTH_combined}
\end{figure}

We have searched the \(\alpha^\prime \times m_\chi\) parameter space following the procedure described in Section~\ref{sec:rotation}, for all galaxies in our sample. The resulting minimized statistic $\chi^2_\mathrm{tot}$ is given, then, by the sum over individual galaxies $\chi^2_\mathrm{tot} = \sum_{i=1}^N \chi^2_{\rm{red},i}$, where \(N\) is the ensemble size. We work with \(\chi^2_{\rm{red}}\) instead of \(\chi^2\) to account for the fact that the number of observations is different for each galaxy. The procedure is executed for three different groups of galaxies: SPARC, LITTLE THINGS, and both. Figures~\ref{fig:SPARC_combined},~\ref{fig:THINGS_combined}, and~\ref{fig:BOTH_combined} show the obtained maps of $\chi^2_\mathrm{tot}$ in the \(\alpha^\prime \times m_\chi\) for each of the three groups.

The behaviour of the minimum of $\chi^2_\mathrm{tot}$ shows the relation between \(m_\chi\) and \(\alpha^\prime\). Through this relation, it is possible to estimate the value of \(\sigma/m_\chi\) for the annihilation process. We can write an equation analogous to \ref{eq:xsec}  

\begin{equation}
    \frac{\sigma}{m_\chi} = \frac{\pi}{v_0}\left(1 - \frac{r_m^2}{4}\right)\frac{\alpha^{\prime^2}}{m^3_\chi},
\end{equation}
where \(v_0\) is a reference velocity for DM in a galactic halo and $r_m\equiv m_\phi /m_\chi$. Writing \(\alpha^\prime\) as a function of \(m_\chi\) gives

\begin{equation}
    \alpha^\prime = \beta \left(\frac{\sigma}{m_\chi}\right)^{1/2} m^{3/2}_\chi,
    \label{eq:function_alpha_m}
\end{equation}
with \(\beta = [v_0/\pi(1-r^2_m/4)]^{1/2}\). The blue line in figures \ref{fig:SPARC_combined}, \ref{fig:THINGS_combined} and \ref{fig:BOTH_combined} shows the fit of the function \ref{eq:function_alpha_m}, where we assumed \(v_0 = 100\) km/s, a value characteristic of DM in a Milky-Way-like galaxy.  

The combined fit performed with LITTLE THINGS data reaches a minimum region more smoothly. This behavior could be explained by the homogeneity of galaxy morphologies in the LITTLE THINGS catalog, which is not true for SPARC. Also, LITTLE THINGS sample has, in general, more data points with large uncertainties for the rotation velocity curves when compared to SPARC. The result for the merge of both catalogs is closer to the one that considers SPARC alone, since the bumps of high-valued $\chi^2_\mathrm{tot}$ in SPARC fit dominate over the lower values of LITTLE THINGS. 

The results for \(\sigma / m_\chi\) are slightly small if compared to the SIDM scattering cross-sections, although for different choices of \(v_0\), they are comparable in order of magnitude with the ones found in Ref.~\citep{Kaplinghat:2015aga}. For example, a lower DM velocity, characteristic of small systems, \(v_0 = 20 \rm{km}/\rm{s}\) gives \(\sigma/m_\chi = (0.533 \pm 0.032)  \rm{cm}^2/g\) considering the LITTLE THINGS sample. The fits show that OADM successfully turned cusp-type halos into cored ones for all the galaxies in our sample, and the results do not depend on the mass of the mediator \(m_\phi\), in accordance with Ref.~\cite{Cline:2020gon}.

\section{Conclusion}
\label{sec:conclusion}

In this paper, we used observational data from galaxy rotation curves to probe the Oscillating Asymmetric Dark Matter (OADM) model with a fermionic dark matter particle coupling to a complex scalar. In this model, a small Majorana mass term violates the conservation of dark matter particle number, leading to oscillations between dark matter and its antiparticle. After freeze-out, these oscillations recouple during the time of structure formation and may transform cusp-type halo density profiles into cored profiles, offering a solution to the core/cusp problem.

We analyzed data from two different rotation curve catalogs: SPARC and LITTLE THINGS. The former includes a more diverse sample in terms of galaxy morphology, while the latter focuses exclusively on dwarf galaxies. The observed rotation curves were compared to those predicted by the OADM model after evolving an initial NFW density profile over \(  10 \ \text{Gyr} \) using an analysis based on solving the Boltzmann equations for the dark matter number density from an initial NFW profile. The fit results demonstrate that the cusp-type profiles were effectively transformed into a cored profiles for all the galaxies in our sample. Although the individual fits for each galaxy return varying preferred values of \(\alpha^{\prime}\) and \(m_\chi\), the combined fits show that it is possible to successfully find values that work well for all galaxies together.

A complementary analysis using N-body simulations could strengthen our results since it has been shown that their results are close to the one obtained by the quantum Boltzmann approach~\citep{Cline:2020gon}. The simulations could, for example, explore scenarios where the DM elastic self-scattering cross section is sizable to the annihilation cross section, the latter of which we have assumed to be dominant. We defer this study for future work. Finally, there remains much phenomenology to be explored with ADM oscillations, which could lead to potential indirect detection signals from the annihilations \citep{Tulin:2012re,Petraki:2013wwa}

\section*{Acknowledgments}
Authors are supported by the São Paulo Research Foundation (FAPESP) through grants number 2021/01089-1 and 2022/16842-0. VdS is supported by CNPq through grant number 308837/2023-1. GG acknowledges support from Universidad
Nacional de Ingenieria funding Grant FC-PFR-42-2024. The authors acknowledge the National Laboratory for Scientific Computing (LNCC/MCTI,  Brazil) for providing HPC resources for the SDumont supercomputer (http://sdumont.lncc.br).\newline

This version of the article has been
accepted for publication, after peer review (when applicable) but is not the Version of Record and does not reflect post-acceptance improvements, or any corrections. The Version of Record is available online at: http://dx.doi.org/10.1140/epjc/s10052-025-13875-x.

\newpage

\onecolumn

\begin{table}[ht]
\caption{\label{table:Table1} The best-fit parameters \(r_s\) and \(\rho_s\) corresponding to the minimum \(\chi^2_{\rm{red}}\) in the \(\alpha^\prime \times m_\chi\) grid, for each galaxy.}
\begin{tabular}{cccccc}
\hline
Catalog/Galaxy            & \(\alpha^\prime\) & \(m_\chi\) [MeV] & \(\rho_s\) [\(\times 10^{-2} M_\odot/\rm{pc}^3\)]            & \(r_s\) [kpc]   & \(\chi^2_{\text{red}}\) \\ \hline \hline
\multicolumn{6}{c} {SPARC} \\ \hline
DDO154      & 0.152             & 448.979              & \(5.28 \pm 1.12\) & \(2.15\pm0.21\)   & 0.9                     \\
DDO170      & 0.229             & 1000.0               & \(0.85\pm0.24\)   & \(5.95\pm0.93\)   & 1.1                     \\
ESO444-G084 & 0.022             & 632.653              & \(0.30 \pm 0.08\) & \(11.35\pm2.38\)  & 0.1                     \\
F563-V2     & 0.264             & 889.796              & \(33.85\pm5.54\) & \(1.60 \pm 0.13\) & 0.1                     \\
F565-V2     & 0.016             & 100.0                & \(0.76\pm0.11\)   & \(9.51\pm0.85\)   & 0.1                     \\
F568-V1     & 0.069             & 375.510              & \(29.63\pm13.24\)   & \(1.92\pm0.35\)   & 0.1                     \\
NGC3109     & 0.199             & 485.741              & \(0.10\pm0.01\)  & \(2.17\pm0.53\)   & 0.2                     \\
UGC00128    & 0.010             & 853.061              & \(0.41\pm0.03\)   & \(16.39\pm0.84\)  & 2.8                     \\
UGC00731    & 0.199             & 908.163              & \(2.60\pm0.57\)   & \(3.82\pm0.42\)   & 0.3                     \\
UGC00891    & 0.069            & 559.184              & \(0.11\pm0.06\)   & \(27.25\pm13.89\) & 0.9                     \\
UGC05716    & 0.010             & 283.673              & \(0.61\pm0.08\)   & \(7.85\pm0.64\)   & 1.9                     \\
UGC05764    & 0.039             & 283.673              & \(70.73\pm17.83\)   & \(0.77\pm0.09\)   & 3.0                     \\
UGC05829    & 0.010             & 118.367              & \(0.24\pm0.01\)   & \(13.56\pm0.57\)  & 0.1                     \\
UGC05918    & 0.016             & 173.469              & \(0.79\pm0.47\)   & \(6.05\pm2.61\)    & 0.6                     \\
UGC06667    & 0.282             & 889.796              & \(7.85\pm0.85\)   & \(3.11\pm0.17\)   & 0.1                     \\
UGC07608    & 0.205             & 761.224              & \(0.75\pm0.25\)   & \(7.72\pm1.79\)   & 0.1                     \\
UGCA442     & 0.259             & 742.857              & \(1.81\pm0.60\)   & \(4.06\pm0.70\)  & 0.6                     \\
UGCA444     & 0.051             & 320.408              & \(1.02\pm0.48\)   & \(3.82\pm1.12\)   & 0.2                     \\ \hline
\multicolumn{6}{c} {LITTLE THINGS} \\ \hline
CVnidwa    & 0.294             & 467.347              & \(16.38\pm8.95\)   & \(2.26\pm0.31\)   & 1.7                    \\
DDO43      & 0.270             & 761.224              & \(11.66\pm5.45\)   & \(1.10\pm0.21\)   & 0.5                     \\
DDO47      & 0.300             & 595.918              & \(3.96\pm0.35\)   & \(4.26\pm0.20\)   & 4.9                     \\
DDO52      & 0.247             & 889.796              & \(3.52\pm1.09\)   & \(3.32\pm0.53\)   & 0.1                     \\
DDO53      & 0.193             & 614.286              & \(13.85\pm6.59\)   & \(1.19\pm0.27\)   & 0.7                     \\
DDO70      & 0.010             & 155.102              & \(0.46\pm0.16\)   & \(8.62\pm2.45\)   & 1.4                     \\
DDO87      & 0.294             & 761.224              & \(0.22\pm0.05\)   & \(13.59\pm1.97\)  & 0.2                     \\
DDO126     & 0.176             & 504.082              & \(9.02\pm1.35\)   & \(1.41\pm0.07\)   & 0.3                     \\
DDO133     & 0.075             & 448.979              & \(0.06\pm1.35\)   & \(1.90\pm0.22\)   & 0.8                     \\
DDO154     & 0.028             & 228.571              & \(2.23\pm0.13\)   & \(3.02\pm0.10\)   & 0.1                     \\
DDO168     & 0.140             & 540.816              & \(13.30\pm7.02\)   & \(2.90\pm0.39\) & 2.1                     \\
DDO210     & 0.028             & 283.673              & \(1..53\pm0.36\)   & \(1.78\pm0.54\)   & 0.3                     \\
DDO216     & 0.010             & 283.673              & \(2.12\pm1.26\)   & \(0.97\pm0.37\)   & 0.2                     \\
F564-V3    & 0.093            & 448.979              & \(28.83\pm14.53\)   & \(0.72\pm0.18\)   & 1.6                     \\
HARO29     & 0.045             & 926.531              & \(4.64\pm1.29\)   & \(1.46\pm0.21\)   & 0.3                     \\
HARO36     & 0.229             & 724.490              & \(0.37\pm0.20\)   & \(18.80\pm7.54\)  & 0.2                     \\
NGC1569    & 0.193             & 577.551              & \(2.74\pm0.33\)   & \(3.40\pm0.39\)   & 0.1                     \\
NGC2366    & 0.069             & 338.775              & \(13.27\pm0.87\)   & \(1.58\pm0.04\)   & 0.1                     \\
UGC8508    & 0.111             & 522.449              & \(0.75\pm0.03\)   & \(13.72\pm0.47\)  & 0.2                     \\
WLM        & 0.241             & 889.796              & \(76.32\pm18.56\)   & \(0.45\pm0.04\)   & 1.5                     \\ \hline
\end{tabular}
\end{table}

\clearpage

\twocolumn


\bibliography{sn-bibliography}

\newpage

\end{document}